\documentclass[12pt]{iopart}
\usepackage{graphicx}% Include figure files
\usepackage{epstopdf}
\begin{document}
\maketitle
%\title[Shell model for $^{71-77}$Ga]
%{High-spin structure of odd $^{71-77}$Ga 
%isotopes with shell model}
\title[Shell model for $^{71-78}$Ga]
{Structure of $^{71-78}$Ga 
isotopes in $f_{5/2}pg_{9/2}$ and $fpg_{9/2}$ spaces}
%{Structure of $^{71-78}$Ga 
%isotopes with shell model}
\vspace{4mm}
\author{P.C. Srivastava  \footnote{Present address: Instituto de Ciencias Nucleares, Universidad Nacional Aut\'onoma de
M\'exico, 04510 M\'exico, D.F., M\'exico, praveen.srivastava@nucleares.unam.mx}}
\address{Physical Research Laboratory, Ahmedabad 380 009, India}
%%\date{\today}
%\ead{praveen@prl.res.in}

\begin{abstract}

We have performed comprehensive set of shell model calculations for Ga isotopes including
high-spin states with three different effective interactions. 
This work will add more information
in the earlier work by 
Cheal {\it et al.}
for odd-even Ga isotopes [ Phys. Rev. Lett. 104, 252502
(2010)] and Man\'e {\it et al.} for odd-odd Ga isotopes
[ Phys. Rev. C 84, 024303
(2011)], where only few excited states are studied in $f_{5/2}pg_{9/2}$ space.
For lighter isotopes $fpg$ interaction is  better and for heavier isotopes jj44b is
quantitatively better than JUN45. These results show that limitation of existing interactions
and calling for further improvements to 
predict nuclear structure properties of Ga isotopes.

\end{abstract}

\vspace{3mm}
%\pacs{21.60.Cs, 27.50.+e}  
%\maketitle
\section {Introduction}
 \label{s_intro} 

\begin{figure}[h!]
\begin{center}
\resizebox{90mm}{!}{\includegraphics{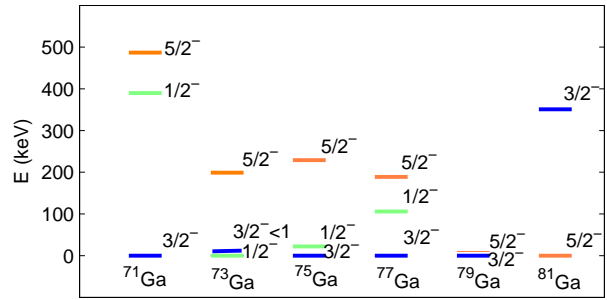}}
\caption{Experimental low-energy systematics of odd-A gallium
isotopes \cite{Stefanescu09,Diriken10,Cheal10}.} 
\label{f_gaintro}
\end{center}
\end{figure}

The structural changes between $N=40$ and $N=50$  of
gallium isotopes recently attracted much
experimental  attention. At Argonne National Laboratory,
Stefanescu {\it et. al.} \cite{Stefanescu09} have populated odd-A $^{71-77}$Ga
isotopes in deep-inelastic reactions.
More recently, in the Coulomb excitation experiment at
\textsc{rex-isolde}, the existence of a $1/2^-$, $3/2^-$ 
ground-state doublet has been proposed  in $^{73}$Ga
\cite{Diriken10}. For odd-even Ga isotopes nuclear spins
and moments has  been reported in \cite{Cheal10}.
Recently ground-state spins and moments of 
 $^{72,74,76,78}$Ga isotopes using laser spectroscopy has been reported in \cite{Mane11}.
The evolution of the  $1/2^-$ and $5/2^-$ levels in odd-A
gallium  isotopes are shown in Fig. \ref{f_gaintro}. Except for 
$^{73}$Ga and
$^{81}$Ga all have ground state (g.s.) $3/2^-$. The first $1/2^-$
reaches minimum in $^{73}$Ga where it becomes the ground state
and the first $5/2^-$ start decreasing with $N=40$ onwards
and it becomes the ground state in  $^{81}$Ga. This figure also
demonstrate abrupt changes of structure from $N=40$ to
$N=42$. 
%Similarly in Cu isotopes rapid reduction in the energy of first $5/2^-$ state has been observed as the filling of neutrons started in the $\nu g_{9/2}$ orbital \cite{Ste08}. In case of Cu, $5/2^-$ is a ground state at $N=46$ and in Ga at $N=50$. 

Following our recent shell-model (SM) studies for
neutron-rich even isotopes of Fe
\cite{Sri09}, odd-odd Mn isotopes \cite{Srib}, and
odd-mass $^{61,63,65}$Co isotopes \cite{Sric}, 
in this paper we report large scale  shell model
calculations for $^{71-78}$Ga isotopes.
 Earlier shell model calculation using pairing plus
quadrupole-quadrupole interaction for $^{75,77,79}$Ga isotopes have been
reported by Yoshinaga {\it et. al.,} \cite{Yoshinaga08}. They
 consider  only low-lying negative-parity states in the analysis.
 The aim of present study is to
analyze recently accumulated experimental data 
which includes both positive and negative high-spin states
on neutron-rich Ga isotopes. Further, this work will add more information
in the earlier work for odd-even Ga isotopes by Cheal {\it et al.} \cite{Cheal10}, 
where only few excited negative-parity states ($< 1$ MeV) are studied in $f_{5/2}pg_{9/2}$ space. 
Present work also include further theoretical development which is proposed 
 in \cite{Cheal10}, by including $f_{7/2}$ orbital in the model space.

\section{ Details of Calculation}

The present shell model calculations have been carried out
in $f_{5/2}pg_{9/2}$ and $fpg_{9/2}$ spaces. In the first set,
calculations have been performed 
with two recently derived effective shell model interactions, JUN45
and jj44b, that have been proposed for 1$p_{3/2}$,
0$f_{5/2}$, 1$p_{1/2}$ and 0$g_{9/2}$ single-particle
orbits. JUN45 which
is recently develop by Honma {\it et al.} \cite{Honma09},
this realistic interaction based on the Bonn-C potential
with fitting 400 experimental binding and excitation
energy data with mass numbers A=63$\sim$96. Since the
present model space is not sufficient to describe
collectivity in these region thus data have been not used
while fitting in the middle of the shell along $N=Z$ line.
For JUN45, the data mostly fitted to develop this
interaction closure to $N=50$, thus this interaction is
very suitable for shell-model study of neutron rich Ga
isotopes as $N \sim 50 $. Although this interaction is not
successful to explain data for Ni and Cu isotopes possibly
due to missing  0$f_{7/2}$ orbit in the present model
space.  The jj44b interaction due to  Brown {\it et al.}
\cite{brown} was developed by fitting 600 binding energies
and excitation energies with $Z=28-30$ and $N=48-50$.
Instead of 45 as in JUN45, here 30 linear combinations of
good $J-T$ two-body matrix elements varied, with the rms
deviation of about 250 keV from experiment.
For JUN45 interaction the single-particle energies
are taken to be -9.8280, -8.7087, -7.8388, and -6.2617 MeV
for the $p_{3/2}$, $f_{5/2}$, $p_{1/2}$ and $g_{9/2}$
orbit, respectively. Similarly for jj44b interaction the
single-particle energies are taken to be -9.6566, -9.2859,
-8.2695, and -5.8944 MeV for the $p_{3/2}$, $f_{5/2}$,
$p_{1/2}$ and $g_{9/2}$ orbit, respectively.

%\begin{figure}[h!]
%\begin{center}
%\resizebox{90mm}{!}{\includegraphics{monopole.eps}}
%\caption{Effective single-aparticle energies of proton orbits for Cu 
%isotopes for JUN45 and jj44b interaction.} 
%\label{f_mono}
%\end{center}
%\end{figure}

In the second set of calculations for the $fpg_{9/2}$ space, 
with $^{48}$Ca core ( a $^{40}$Ca core, with eight neutrons were frozen in the $\nu f_{7/2}$ orbital), an interaction
reported  by Sorlin {\it et al}. in \cite{Sorlin02} has been
employed.
This interaction, called $fpg$ interaction, was built
using $fp$ two-body matrix elements (TBME) from
 \cite{Pov01} and $rg$ TBME from \cite{Nowacki96}. The
remaining $f_{7/2} g_{9/2}$  TBME are taken from  \cite{Kahana69}.
The single-particle energies
are taken to be 0.0, 2.0, 4.0, 6.5 and 9.0 MeV
for the $f_{7/2}$, $p_{3/2}$, $p_{1/2}$, $f_{5/2}$,  and $g_{9/2}$
orbit, respectively.
As here the
dimensions of the matrices become very large, a truncation has
been imposed. We used a truncation by allowing up to a total of
three particle excitations from the $f_{7/2}$ orbital to the
upper  ${fp}$ orbitals for protons and from the upper $fp$
orbitals to the $g_{9/2}$ orbital for neutrons.

%In Fig. \ref{f_mono}, we compared the effective single-particle energy
%of  the proton  orbit for Cu isotopes in JUN45 and jj44b
%interaction.  Both the interactions show a rapid decrease
%in $f_{5/2}$ proton single-particle energy relative to
%$p_{3/2}$ as neutrons start filling in $g_{9/2}$ orbit and
%it becomes lower than $p_{3/2}$  for $N>48$.

 All the SM calculations are performed
on the 20-node cluster computer at PRL using the
code \textsc{antoine} ~\cite{Antoine}. 

\section{Results for odd-even Ga isotopes }

%\subsection{Excitation energies and comparison with the data}

%\newpage

%\includegraphics[width=12.5in]{71Gan.eps}\\
%\resizebox{170mm}{130mm}{\includegraphics{71Gann.eps}}

\subsection{{\bf $^{71}$Ga :}}

\begin{figure}
\begin{center}
\includegraphics[scale=0.45]{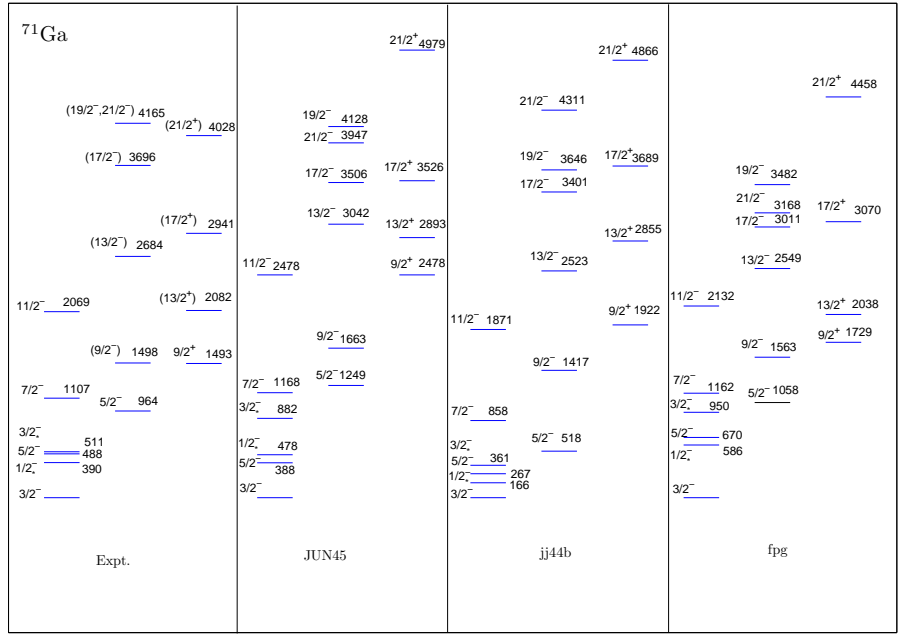}
\caption{Experimental data for $^{71}$Ga compared
with shell-model results. The levels marked with an asterisk (*) 
have been also experimentally found, but not form a band.}
\label{f_71Ga}
\end{center}
\end{figure}

Stefanescu {\it et al.}
\cite{Stefanescu09} assigned positive parity states at 2082, 2941, and 4028
keV,  for 13/2$^+$, 17/2$^+$, and 21/2$^+$ respectively. In the same
experiment the negative  states 9/2$^-$, 13/2$^-$, 17/2$^-$, and
(19/2$^-$, 21/2$^-$) are proposed at 1498, 2684, 3696, and 4165 keV,
respectively. In this paper they argue that these sequence of levels
arises due to coupling of $\pi f_{5/2}$ and $\nu g_{9/2}$.  In Fig. \ref{f_71Ga}, experimental data  for $^{71}$Ga
are compared with the calculated energy levels using JUN45, jj44b and $fpg$
interactions. The band built on
3/2$^-$, is reproduced correctly by JUN45 interaction, while levels
are compress with jj44b interaction. 
These results show that the yrast sequence of levels 
$3/2^-$--$5/2^-$--$7/2^-$--$11/2^-$ is connected with strong $E2$ transitions.
 The $B(E2)$
values for 5/2$^-$$\rightarrow$3/2$^-$, 7/2$^-$$\rightarrow$5/2$^-$,
11/2$^-$$\rightarrow$7/2$^-$, are 6, 222 and 223 e$^2$fm$^4$
respectively with JUN45 and corresponding value for jj44b
are 5, 209 and 383 e$^2$fm$^4$. 
Similarly, a band built on
5/2$^-$, with JUN45 interaction is slightly higher, while jj44b
interaction predicted compress states. The band built on 9/2$^+$, 
with the JUN45 interaction, predicted
9/2$^+$, 13/2$^+$, 17/2$^+$, 21/2$^+$ are at 2478, 2893,
3526 and 4979 keV, while its corresponding experimental 
values are 1493, 2082, 2941 and 4028 keV respectively. The
calculated positive parity states around 1 MeV higher than
experimental data with JUN45. The predicted results show  an
inadequacy of this interaction to explain positive parity
states. The jj44b interaction predicts 
9/2$^+$, 13/2$^+$, 17/2$^+$, 21/2$^+$ at 1922, 2855,
3689 and 4866 keV respectively, which is better than JUN45
interaction. 
The calculated $B(E2)$ values are shown in Table \ref{t_be2}. The JUN45 interaction predicts, leading configuration of $\pi(p_{3/2}^3)$$\otimes$$\nu(p^4_{3/2}f^6_{5/2}p^2_{1/2})$ for the ground state with probability of 29.2\%. The calculated occupancy of g$_{9/2}$ orbital is
0.53, 0.83 by  JUN45 and jj44b interaction respectively.
States with $I$= 5/2$^-$, 7/2$^-$, 9/2$^-$, 11/2$^-$, 13/2$^-$
are dominated with $\pi(p_{3/2}^2f_{5/2}^1)$ $\otimes$ $\nu ( p_{3/2}^4f_{5/2}^4p_{1/2}^2g_{9/2}^2)$ configurations by $fpg$ interaction. While for this interaction positive parity states with $I$= 9/2$^+$, 13/2$^+$, 17/2$^+$, 21/2$^+$
are dominated with $\pi(p_{3/2}f_{5/2})^3$ $\otimes$ $\nu ( p_{3/2}^4f_{5/2}^5p_{1/2}^2g_{9/2}^1)$ configurations. The 13/2$^+$ predicted by $fpg$ interaction is within 50 keV with experimental data. 
The $fpg$ interaction predict better result for excitation energies in comparison to JUN45  and jj44b interactions.

% Here if we allowed maximum five particle excitations instead of three then levels are more compressed. 

%\newpage
\begin{figure}
\begin{center}
\includegraphics[scale=0.45]{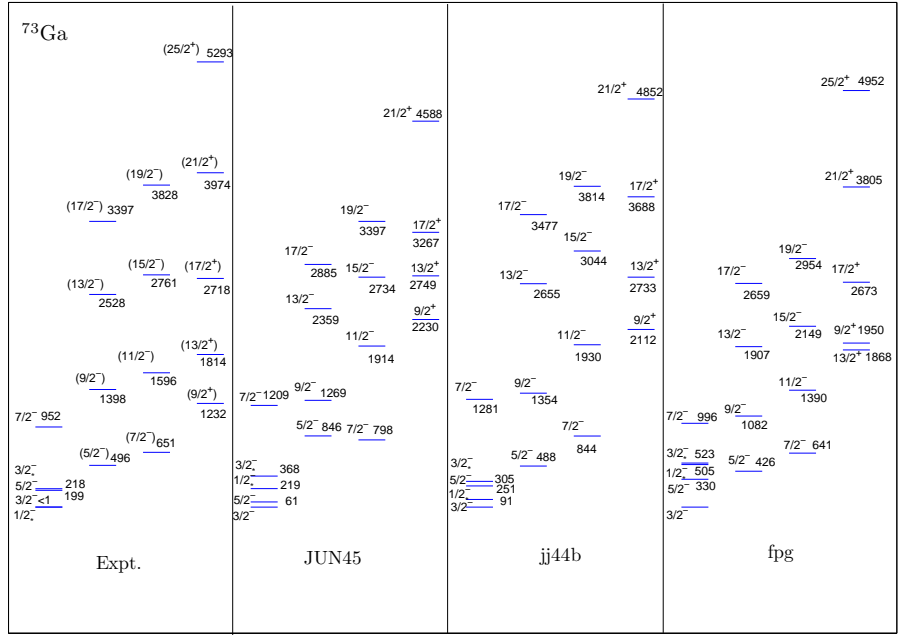}
%\resizebox{170mm}{130mm}{\includegraphics{73Gan.eps}}
\caption{ Experimental data for $^{73}$Ga compared
with shell-model results. The levels marked with an asterisk (*) 
have been also experimentally found, but not form a band.}
\label{f_73Ga}
\end{center}
\end{figure}

\subsection{ {\bf $^{73}$Ga :}}

 Cheal {\it et al.} \cite{Cheal10}
assigned 1/2$^-$ as a
ground state, in another experiment recently Diriken {\it et al.} \cite{Diriken10} shown an evidence for a 1/2$^-$, 3/2$^-$
doublet near the ground state, with an excitation energy of the
$\sim$ 0.8 keV for 3/2$^-$. In Fig. \ref{f_73Ga}, experimental data  for
$^{73}$Ga are compared with the calculated energy levels using JUN45, jj44b and $fpg$
interactions. In  contrast to ground state doublet
predicted in recent experiment \cite{Diriken10}, the JUN45, jj44b and $fpg$ 
interactions predict 1/2$^-$ state at 219, 91 and 505  keV respectively. Indeed,
JUN45 predict too high 1/2$^-$ for $^{73}$Ga. Similar trend has been also observed
with JUN45 interaction for Cu isotopes where 5/2$^-$ is too high below $N=40$ and 
1/2$^-$ is too high above $N=40$.

The band built on 3/2$^-$, reproduced correctly by JUN45 and jj44b  interaction.
The band built on 5/2$^-$, is more correctly reproduced by jj44b interaction.
The $B(E2)$ values for 9/2$^-$$\rightarrow$5/2$^-$, 13/2$^-$$\rightarrow$9/2$^-$,
17/2$^-$$\rightarrow$13/2$^-$, are 42, 194 and 4 e$^2$fm$^4$ respectively with
JUN45 and corresponding value for jj44b are 81, 270 and 161 e$^2$fm$^4$.  The
results of band built on 7/2$^-$, is better for jj44b in comparison to JUN45.
The band built on 9/2$^+$, with the JUN45 interaction, predicted 9/2$^+$,
13/2$^+$, 17/2$^+$, 21/2$^+$ are at 2230, 2749, 3267 and 4588 keV, while its
corresponding experimental  values are 1232, 1814, 2718 and 3974 keV
respectively. The
occupancy of $g_{9/2}$ orbital is predicted by  JUN45 and jj44b 
interaction as 0.92 and 0.83 respectively.
 With $fpg$ interaction 13/2$_1^+$
become lower than 9/2$_1^+$. 
The calculated $B(E2)$ values obtained with jj44b
give reasonable agreement with experimental data.
The experimental excitation energy of 9/2$_1^+$ decreases from $^{71}$Ga to $^{73}$Ga and again starts rising from $^{75}$Ga 
onwards. Indeed, intruder $\pi g_{9/2}$ leading to onset of 
deformation around neutron mid-shell at $N=42$.
Only JUN45 interaction predict this trend from $^{71}$Ga to $^{73}$Ga. 
States with $I$= 7/2$^-$, 9/2$^-$, 11/2$^-$, 13/2$^-$, 15/2$^-$, 17/2$^-$, 19/2$^-$, 21/2$^-$ 
are dominated with $\pi(p_{3/2}f_{5/2})^3$ $\otimes$ $\nu ( p_{3/2}^4f_{5/2}^4p_{1/2}^2g_{9/2}^4)$ configurations by jj44b interaction for $^{73}$Ga.
The overall results of jj44b interaction are better than JUN45 and $fpg$.

%\newpage
\begin{figure}
\begin{center}
\includegraphics[scale=0.42]{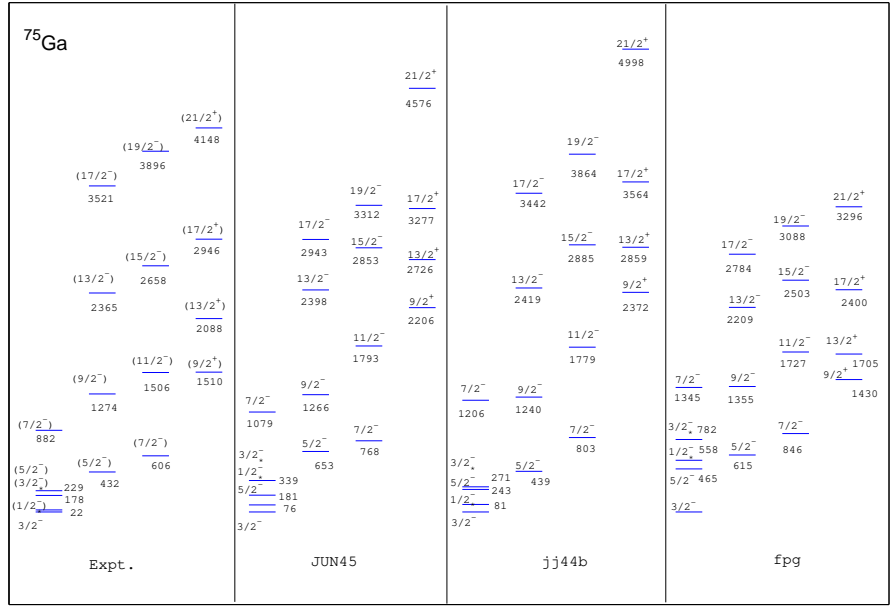}
%\resizebox{170mm}{130mm}{\includegraphics{75Gan.eps}}
\caption{Experimental data for $^{75}$Ga compared
with shell-model results. The levels marked with an asterisk (*) 
have been also experimentally found, but not form a band.}
\label{f_75Ga}
\end{center}
\end{figure}

\subsection{ {\bf $^{75}$Ga :} }

 In Fig.\ref{f_75Ga}, experimental data  for $^{75}$Ga are compared with
the calculated energy levels with JUN45, jj44b and $fpg$
interactions. The band built on 3/2$^-$, reproduced correctly by JUN45 and jj44b 
interaction. The band built on 5/2$^-$, is more correctly reproduced
by jj44b interaction. The results of band built on 7/2$^-$ is reasonable
with jj44b. The $B(E2)$ values for
11/2$^-$$\rightarrow$7/2$^-$, 15/2$^-$$\rightarrow$11/2$^-$,
19/2$^-$$\rightarrow$15/2$^-$, are 278, 109 and 92 e$^2$fm$^4$
respectively with JUN45 and corresponding values for jj44b are 388,
368 and 281 e$^2$fm$^4$. The band built on 9/2$^+$, with the JUN45
interaction, predicted 9/2$^+$, 13/2$^+$, 17/2$^+$, 21/2$^+$ are at
2206, 2726, 3277 and 4576 keV, while its corresponding experimental 
values are 1510, 2088, 2946 and 4148 keV respectively. The ground state having
configuration   $\pi(p_{3/2}f_{5/2}^2)$ with probability 14.9 and
22.4\% for JUN45 and jj44b interaction.
The results of $fpg$ interaction
are more compressed. 
States with $I$= 7/2$^-$, 9/2$^-$, 11/2$^-$, 13/2$^-$, 15/2$^-$, 17/2$^-$, 19/2$^-$, 21/2$^-$ 
are dominated with $\pi(p_{3/2}f_{5/2})^3$ $\otimes$ $\nu ( p_{3/2}^4f_{5/2}^4p_{1/2}^2g_{9/2}^6)$ configurations by jj44b interaction for $^{75}$Ga.
The first $9/2^+$ has $\pi (f_{5/2}^2g_{9/2}^1$) $\otimes$ $\nu ( p_{3/2}^4f_{5/2}^4p_{1/2}^2g_{9/2}^6)$
configuration with jj44b for $^{75}$Ga.
With jj44b the occupancy of
$g_{9/2}$ orbital is 0.49.
The results of jj44b interaction are better
than JUN45 and $fpg$.

%The band head energies for 5/2$^-$
%and 7/2$^-$ band predicted by $fpg$ is higher than experimental value.
%\newpage
\begin{figure}
\begin{center}
\includegraphics[scale=0.40]{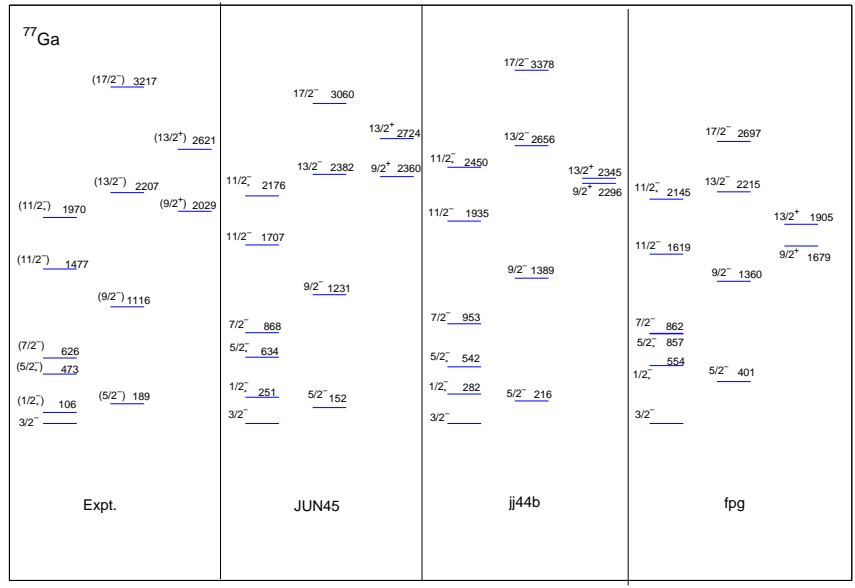}
%\resizebox{170mm}{130mm}{\includegraphics{77Gan.eps}}
\caption{Experimental data for $^{77}$Ga compared
with shell-model results. The levels marked with an asterisk (*) 
have been also experimentally found, but not form a band.}
\label{f_77Ga}
\end{center}
\end{figure}

\subsection{ {\bf $^{77}$Ga :}}

In Fig.\ref{f_77Ga}, experimental data  for $^{77}$Ga are compared with
the calculated energy levels with JUN45, jj44b and $fpg$
interactions. The band built on 3/2$^-$, is
reproduced correctly by JUN45 interaction, while levels are slightly
higher with jj44b interaction. The $B(E2)$ values for
7/2$^-$$\rightarrow$3/2$^-$, 11/2$^-$$\rightarrow$7/2$^-$, are 45 and
265 e$^2$fm$^4$ with JUN45 and corresponding value for jj44b is 38
and 310 e$^2$fm$^4$. 
 The $B(E2)$ values for
9/2$^-$$\rightarrow$5/2$^-$, 13/2$^-$$\rightarrow$9/2$^-$,
17/2$^-$$\rightarrow$13/2$^-$, are 135, 33 and 105 e$^2$fm$^4$
respectively with JUN45 and corresponding value for jj44b are 167, 78
and 101 e$^2$fm$^4$. The band built on 9/2$^+$ is correctly reproduce by JUN45
interaction. The $13/2^+$ is predicted at 2724 keV by JUN45,
while at 2345 keV by jj44b, which is slightly compressed.
The ground state as
$3/2^-$ having configuration 
$\pi(p_{3/2}f_{5/2}^2)$ with probability 35.7 and 24.5\% for JUN45 and
jj44b interaction respectively. 
With jj44b interaction the 1/2$^-$ is high for $^{77}$Ga. Similar
trend has been also found for Cu isotopes, where 1/2$^-$ bit too high in $^{73,75}$Cu.
States with $I$= 7/2$^-$, 9/2$^-$, 11/2$^-$, 13/2$^-$, 15/2$^-$, 17/2$^-$, 19/2$^-$, 21/2$^-$ 
are dominated with $\pi(p_{3/2}f_{5/2})^3$ $\otimes$ $\nu ( p_{3/2}^4f_{5/2}^6p_{1/2}^2g_{9/2}^6)$ configurations by JUN45 interaction for $^{77}$Ga.
The occupancy of g$_{9/2}$ orbital
for first $9/2^+$ state is 0.96 for JUN45, while it is only 0.053 for
jj44b interaction.
The results of JUN45 and jj44b are reasonable.
The positive parity states predicted by $fpg$ are more compressed.

\begin{figure*}
\begin{center}
\includegraphics[scale=0.45]{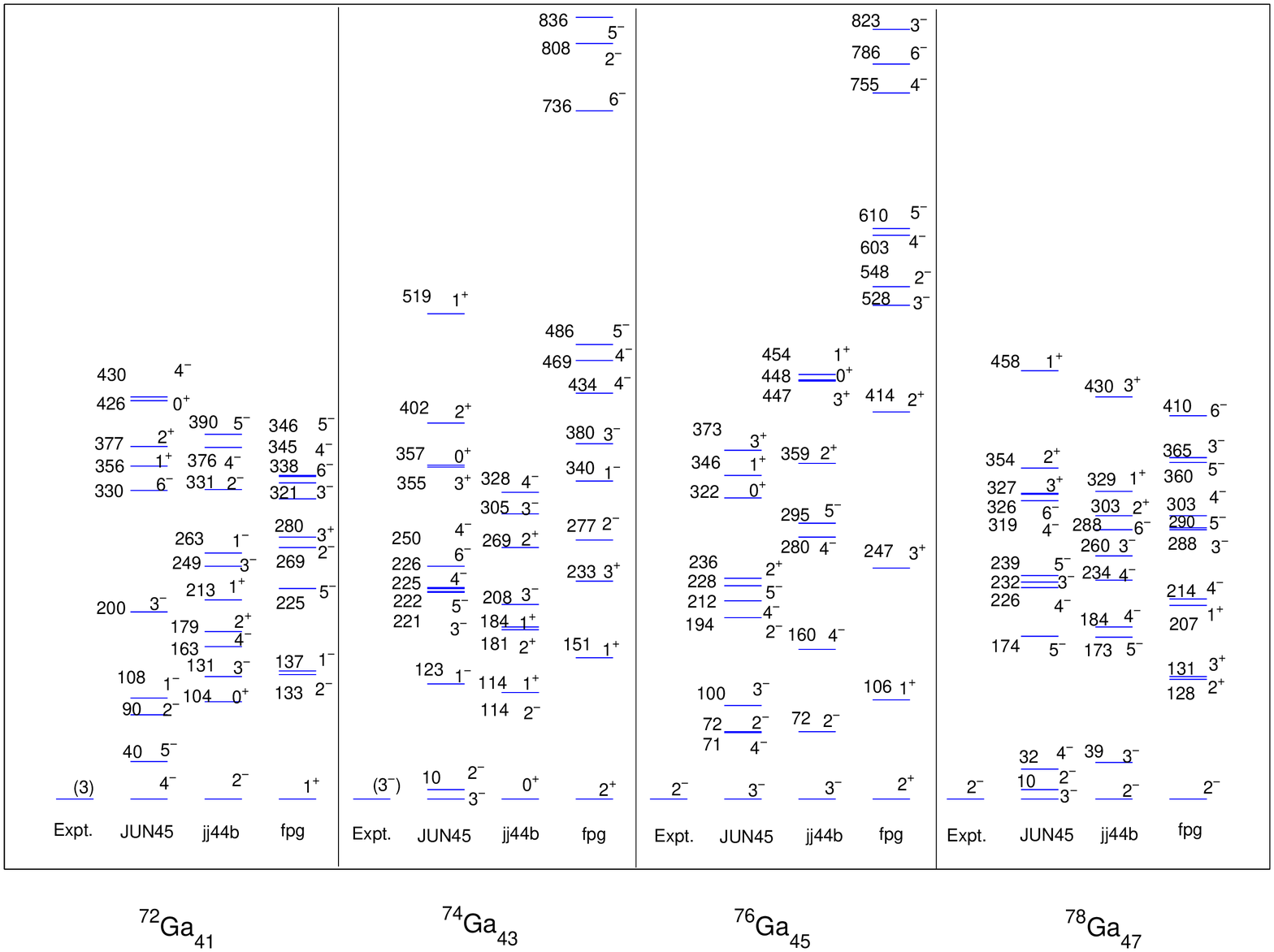}
%\resizebox{170mm}{130mm}{\includegraphics{77Gan.eps}}
\caption{ Calculated levels of $^{72,74,76,78}$Ga compared to experimental levels using three different interactions.}
\label{f_77Ga}
\end{center}
\end{figure*}

\section{Results for odd-odd Ga isotopes }

In Fig. 6 the calculated energy levels for even Ga isotopes with JUN45, jj44b and $fpg$ interactions are compared with experimentally known data. 
Recently ground-state spins and moments for even Ga
isotopes with A=72,74,76,78 has been reported in \cite{Mane11}.
For $^{72}$Ga all the three interactions
predict different g.s. The first 3$^-$ is predicted at 200, 131 and 321 keV by JUN45, jj44b and $fpg$ 
interaction respectively. In $^{74}$Ga the JUN45 interaction predict 3$^-$ as a g.s. which is in agreement with tentative assignment  \cite{Mane11}. While with jj44b and $fpg$ it is more than 200 keV higher.
The jj44b interaction predict gradual drop in 2$^-$ energy from $^{74}$Ga to $^{78}$Ga. This is
reasonable trend because for $^{78}$Ga it is a g.s. which is in agreement with experimental data. 
As we move from $^{72}$Ga onwards, the contribution in g.s. wavefunction for $\pi(f_{5/2}^3)$ 
configuration is start increasing.
There is also similarity between spectra of gallium and copper isotopes such as 2$^-$ is a g.s. in case of $^{72,74}$Cu and $^{76,78}$Ga. The leading proton
configuration in $^{72,74}$Cu is $\pi(f_{5/2})$ which show an agreement with experimental moments.

\begin{table}
\caption{Calculated $B(E2)$ values for some transition for
 $^{71-77}$Ga isotopes with standard effective charges:
 e$_{\rm eff}^\pi$=1.5$e$, e$_{\rm eff}^\nu$=0.5$e$ (the experimental $\gamma$-ray energies
 corresponding to these transitions are also shown).}
\begin{center}
\resizebox{11.5cm}{!}{
\begin{tabular}{c|c|c|c|c|c|c}
\hline
&&&\multicolumn{3}{c}{~~$B(E2)$ (W.u.)}\\
\cline{4-7}
Nucleus & $I_i^\pi \rightarrow I_f^\pi $ & $E_\gamma$ (keV) &~ Expt. &~  JUN45 &~  jj44b &~  $fpg$\\		   
\hline
$^{71}$Ga~~~ & $5/2_2^-$ $\rightarrow$ ~3/2$_1^-$  &~~ 965&~~ 9.0(5) &~ 4.43 &~~ 16.82 &~~ 4.93 \\
%\cline{2-4}
             & $7/2_1^-$ $\rightarrow$ ~3/2$_1^-$  &~~ 1107&~~ 0.8(1) &~ 1.12 &~~  4.34 &~~ 0.59 \\
%\cline{2-4}
             & $7/2_2^-$ $\rightarrow$ 3/2$_1^-$   &~~ 1395&~~ 2.8(4) &~ 1.77 &~~   8.38 &~~ 1.55 \\

\hline
$^{73}$Ga~~~ & $5/2_1^-$ $\rightarrow$ ~1/2$_1^-$   &~~ 199&~~ 11.0(2) &~ 7.58 &~~ 7.36 &~~ 0.17\\
%\cline{2-5}
             & $3/2_2^-$ $\rightarrow$ ~1/2$_1^-$   &~~ 218&~~ 7.5(10) &~  11.62 &~~ 7.92 &~~ 5.50\\
%\cline{2-5}
             & $5/2_2^-$ $\rightarrow$ ~1/2$_1^-$   &~~ 496&~~ 6.5(10) &~  3.65 &~~ 6.38 &~~ 9.65\\
%\cline{2-5}
             & $5/2_3^-$ $\rightarrow$ ~1/2$_1^-$  &~~ 1395&~~ 3.0(7) &~ 0.265 &~~ 2.60 &~~ 8.49\\
%\cline{2-5}
\hline
$^{75}$Ga~~~ & $3/2_2^-$ $\rightarrow$ ~3/2$_1^-$   &~~ 178&~~ - &~ 9.44 &~~ 4.47 &~~ 2.24\\
%\cline{2-5}
             & $5/2_2^-$ $\rightarrow$ ~3/2$_1^-$   &~~ 229&~~ - &~ 0.09 &~~ 0.28 &~~ 2.38\\
%\cline{2-5}
             & $7/2_2^-$ $\rightarrow$ ~3/2$_1^-$   &~~ 606&~~ - &~ 0.34 &~~ 5.09 &~~ 4.13\\
\hline
$^{77}$Ga~~~ & $5/2_1^-$ $\rightarrow$ ~3/2$_1^-$   &~~ 189&~~ - &~ 0.75 &~~ 7.62 &~~ 14.96\\
%\cline{2-5}
             & $3/2_2^-$ $\rightarrow$ ~3/2$_1^-$   &~~ 473&~~ - &~  7.40 &~~ 6.74 &~~ 3.53\\
%\cline{2-5}
             & $7/2_2^-$ $\rightarrow$ ~3/2$_1^-$   &~~ 626&~~ - &~ 2.30&~~ 1.94 &~~ 3.03\\
\hline           
\end{tabular}}
\end{center}
\label{t_be2}
\end{table} 

%\newpage
\begin{table}%[h!]
\caption{ Calculated and experimental\cite{Cheal10, Mane11, Cheal10a} quadrupole moments and magnetic moments of Ga isotopes. For even Ga isotopes these state may not be predicted as ground state by shell-model.}
\begin{center}
\resizebox{16.0cm}{!}{
\begin{tabular}{|c|c|c|c|c|c||c|c|c|c|c|}
\hline
Nucleus & $I$ & $Q_{s,expt}$(eb) & $Q_{s,fpg}$(eb)&  $\mu_{expt}(\mu_N)$ & $\mu_{fpg}(\mu_N)$ & Nucleus & $I$ & $\pi_{\small{SM}}$ & $Q_{s,expt}$(eb) & $Q_{s,fpg}$(eb)\\	   
\hline
$^{71}$Ga& $3/2_1$  & +0.106(3) &~~ +0.166 & +2.56227(2) &~~ +2.198 & $^{72}$Ga& $3_1$ & $-$ & +0.536(29) &~~ +0.017 \\
$^{73}$Ga& $1/2_1$  &   0    &~~ 0  & +0.209(2)    &~~ +0.039          & $^{74}$Ga& $3_1$ & $-$ & +0.549(40)&~~ +0.425 \\ 
$^{75}$Ga& $3/2_1$  & $-$0.285(17) &~~ $-$0.338 &  +1.836(4)    &~~ +1.715  & $^{76}$Ga& $2_1$ & $-$ & +0.329(19) &~~ +0.268\\
$^{77}$Ga& $3/2_1$  & $-$0.208(13) &~~ $-$0.289 &  +2.020(3)    &~~ +1.831  & $^{78}$Ga& $2_1$ & $-$ & +0.327(18) &~~ +0.381\\

$^{79}$Ga& $3/2_1$  & +0.158(10) &~~  $-$0.074 & +1.047(3)  &~~  +1.489  & $^{80}$Ga& $6_1$ & $-$ & +0.478(27) &~~ +0.568\\
$^{81}$Ga& $5/2_1$ &  $-$0.048(8) &~~ +0.042  & +1.747(5) &~~ +1.644   & $^{80}$Ga& $3_1$ & $-$ & +0.375(21) &~~ +0.394\\

\hline           
\end{tabular}}
\end{center}
\label{t_quad}
\end{table} 

\section{ {\bf Electromagnetic properties}}
%\section{ {\bf Quadrupole moments and magnetic moments}}
As proposed in \cite{Cheal10} possibility of proton excitations across
Z=28 may play important role, by including the proton $f_{7/2}$
orbital in the model space. 
We have calculated 
static quadrupole moments with effective charges 
$e_p^{eff}=1.5e, e_n^{eff}=1.1e$ and magnetic moments with $g_s^{eff}=0.7g_s^{free}$ 
%for the ground state of $^{71,73,75,77}$Ga 
in  ${fpg_{9/2}}$ model space as shown in Table \ref{t_quad}.
The $fpg$ interaction predict correct
sign of quadrupole moments for $^{71,73,75,77}$Ga and results are better than if we use only $f_{5/2}pg_{9/2}$ space
\cite{Cheal10}. 
The change  in sign of
quadrupole moments from $^{71}$Ga, i.e. $N=40$ onwards, demonstrate
a changing of shell structure. This is due to ground state of $^{71}$Ga have $\pi(p_{3/2}^3)$ configuration
($\sim$ 28\%) and $^{75,77}$Ga have $\pi p_{3/2}^1(f_{5/2}p_{1/2})^2$ configuration. These configurations reveal change of sign of quadrupole moments because filling of higher orbital started. Thus below $N=42$, a hole configuration ($\pi(p_{3/2}^3)$) has
a positive quadrupole moment and above $N=42$, 
a particle configuration ($\pi p_{3/2}^1(f_{5/2}p_{1/2})^2$) has a negative quadrupole moment. 
The $fpg$ interaction predict correct sign of magnetic moment for $^{73}$Ga while JUN45 and jj44b gives
negative sign. This support ground state as 1/2$^-$ for $^{73}$Ga.
We have
also reported quadrupole moments for $^{72,74,76,78,80}$Ga isotopes with $fpg$ interaction.
For $^{74}$Ga the $fpg$ give reasonable value of quadrupole moment in comparison to jj44b \cite{Mane11} while it predict too low value for $^{72}$Ga. 

% In case of $^{73}$Ga small contribution
%of wave function $\sim$ 3$\%$  from $\pi(
%f_{7/2}^7p_{3/2}^2f_{5/2}^2)$ configuration. For $^{75}$Ga maximum contribution 45$\%$ with $\pi(p_{3/2}^1f_{5/2}^2)$ configuration and $\sim$ 5$\%$  with $\pi(f_{7/2}^7p_{3/2}^2f_{5/2}^2)$ configuration.

\section{Summary}

In the present work large scale shell model calculations have been performed for
 $^{71-78}$Ga  isotopes in $f_{5/2}pg_{9/2}$ and
$fpg_{9/2}$ model spaces with $^{56}$Ni and $^{48}$Ca core respectively.
For the $f_{5/2}pg_{9/2}$ space, calculations have been performed with JUN45 and
 jj44b interactions and for $fpg_{9/2}$ model space with $fpg$ interaction.
In $fpg_{9/2}$ space, we use a truncation by allowing maximum three particle excitations.
  
Our calculated results
predict high-spin sequences built
on top of the 3/2$^-$, 5/2$^-$ and 9/2$^+$ levels in odd $^{71-77}$Ga.
The results of $fpg$ interaction are better
than JUN45 and jj44b in lighter isotopes.
 While for heavier isotopes, the
results of jj44b interaction are quantitatively better than JUN45.
The results of level energies for heavier Ga isotopes with $fpg$ interaction are more compressed.
The calculated $B(E2)$ values for $^{73}$Ga with jj44b interaction give reasonable agreement with experimental data. 
%The calculated quadrupole moment in $^{71}$Ga is close to experimental data with correct sign
%with $fpg$ interaction. 
We have also pointed out that
the energy level systematics of Ga isotopes are quite similar to that in Cu isotopes.
There is a clear need for suitable interaction for neutron-rich nuclei
in this region for better understanding of nuclear structure.

%The overall results of $fpg$ interaction calling for further theoretical developments for
%suitable interaction for ${fpg_{9/2}}$ model space.
\section*{Acknowledgments}
Thanks are due to V.K.B. Kota, M. Honma, J. Diriken (for $B(E2)$ values), R. Sahu and I. Mehrotra  for useful discussions. I would also like to thank F. Nowacki for supplying $fpg$ interaction matrix elements.

\end{document}